\documentclass[reprint,superscriptaddress,aps,prl,amsmath,amssymb,floatfix,notitlepage]{revtex4-1}
\usepackage{amsthm}
\usepackage{amsmath}
\usepackage{latexsym}
\usepackage{amsfonts}
\usepackage{amssymb}
\usepackage{color}
\usepackage{bbm,dsfont}
\usepackage{graphicx}
\usepackage{hyperref}
\usepackage{tikz}
\usetikzlibrary{arrows,matrix}
\usepackage{placeins}
\usepackage[utf8]{inputenc}
\usepackage{xr}
\newtheorem{proposition?}{Proposition?}

\theoremstyle{definition}

\newcommand{\complex}{\mathbb C} %complex
\newcommand{\hi}{\mathcal{H}} %Hilbert space H
\newcommand{\kb}[2]{|#1\rangle\!\langle#2|} % ketbra
\newcommand{\ket}[1]{|#1\rangle} %ket
\newcommand{\bra}[1]{\langle#1|} %bra
\newcommand{\tr}[1]{\text{tr}\left\{#1\right\}}
\newcommand{\id}{\mathbbm{1}} %identity operator
\newcommand{\diff}{\textrm{d}} %d
\newcommand{\mean}[1]{\mathsf{M}\left[#1\right]}
\newcommand{\Ex}[1]{{\mathsf{E}}\left[ #1 \right]}
\newcommand{\abs}[1]{\vert #1 \vert}
\newcommand{\norm}[1]{\vert\vert #1\vert\vert}

\begin{document}
\title{Diffusive limit of non-Markovian quantum jumps}
\author{Kimmo Luoma}
\email{kimmo.luoma@tu-dresden.de}
\affiliation{Institut f{\"u}r Theoretische Physik, Technische Universit{\"a}t Dresden, 
 D-01062, Dresden, Germany}
\author{Walter T. Strunz}
%\email{walter.strunz@tu-dresden.de}
\affiliation{Institut f{\"u}r Theoretische Physik, Technische Universit{\"a}t Dresden, 
 D-01062, Dresden, Germany}
%\affiliation{Institut f{\"u}r Theoretische Physik, Technische Universit{\"a}t Dresden, 
%D-01062, Dresden, Germany}

%\homepage[]{Your web page}
%\thanks{}
%\altaffiliation{}

\author{Jyrki Piilo}
%\email{jyrki.piilo@utu.fi}
\affiliation{Turku Center for Quantum Physics, Department of Physics and 
Astronomy, University of Turku, FI-20014, Turun Yliopisto, Finland}

%Collaboration name if desired (requires use of superscriptaddress
%option in \documentclass). \noaffiliation is required (may also be
%used with the \author command).
%\collaboration can be followed by \email, \homepage, \thanks as well.
%\collaboration{}
%\email{}
%\noaffiliation

\date{\today}

\begin{abstract}
 We solve two long standing problems for stochastic descriptions of open quantum system dynamics. 
  First, we find  the  classical stochastic processes corresponding to non-Markovian quantum state diffusion and non-Markovian quantum
  jumps in projective Hilbert space.  
  Second, we show  that the diffusive limit of non-Markovian quantum
  jumps can be taken on the projective Hilbert space in such a 
  way that it coincides with non-Markovian quantum state diffusion.
  However, the very same limit taken on the Hilbert space leads to   
  a completely new diffusive unraveling, which we call  non-Markovian quantum diffusion. 
  Further, we expand the applicability of non-Markovian quantum jumps  and non-Markovian
  quantum diffusion by using a kernel smoothing  technique allowing a significant simplification in their use.
  Lastly, we demonstrate the 
  applicability of our results by studying {a driven dissipative two level atom}
  in a non-Markovian regime 
  using all of the three methods. 
\end{abstract}
\maketitle
\paragraph{Introduction.---}
Deriving and solving the equations of motion for 
driven dissipative quantum systems is a notoriously  
hard task, especially in the presence of  quantum memory effects. In this Letter,
we open new avenues to tackle these problems of broad  on-going interest.
Currently, state-of-the-art experiments explore driven dissipative open quantum 
systems~\cite{RevModPhys.80.885},  non-equilibrium phase transitions
in a  Rydberg gas has been observed~\cite{PhysRevA.96.041602}, simulation
of general open system dynamics with trapped ions has been reported~\cite{Barreiro2011,M_ller_2011} -- and 
even the statistical likelihood of a physical process (a statistical arrow of time) 
has been experimentally characterized using superconducting qubit systems~\cite{PhysRevLett.123.020502}.
Similar type of open quantum systems appear also in the context of photosynthesis~\cite{Engel2007,Lee1462} and 
in general in molecular aggregates~\cite{doi:10.1002/aenm.201700236}.

One of the main difficulties in analyzing 
driven open quantum systems has its origin in the lack of a typical time scale, such as 
an energy gap of the system Hamiltonian. One possible 
solution is to try to model the open system and environment dynamics exactly,
as in non-Markovian quantum state diffusion~\cite{PhysRevA.58.1699,PhysRevLett.82.1801}, where 
a stochastic Schrödinger equation describes the dynamics of the open system and the effects
of the environment are contained in the statistical properties of the driving noise. Typically
approximation methods are required to solve the resulting equations of 
motion~\cite{PhysRevA.60.91,PhysRevLett.113.150403,Hartmann2017}. 
This type of 
approach has been successfully used to describe energy~\cite{PhysRevLett.103.058301,Ritschel2011,doi:10.1063/1.4863968} and
charge transport~\cite{doi:10.1063/1.5095578,doi:10.1063/1.4773319} in molecular
aggregates.

Alternatively, starting from a microscopic model an effective time local master equation can 
be derived~\cite{breuer2002theory} and unravelled, for example, with 
non-Markovian quantum jumps~\cite{PhysRevLett.100.180402,PhysRevA.79.062112,H_rk_nen_2010}.
Quantum jump methods  have been used earlier, e.g., to study excitonic energy transport with~\cite{Ai_2014,Tao2016} and without 
driving~\cite{doi:10.1063/1.3259838,doi:10.1063/1.5048058} and even
to understand singlet fission in molecular crystals, which may help to design
more efficient solar panels~\cite{Renaud2015}.   

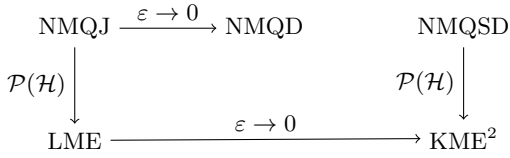
\begin{figure}[h!]
  \begin{tikzpicture}
    \matrix (m) [matrix of math nodes,row sep=3em,column sep=4em,minimum width=2em]
    {
      |[name=nmqj]| \text{NMQJ}&  |[name = nmd]| \text{NMQD} & | [name=nmqsd]| \text{NMQSD}\\
      |[name = lme]| \text{LME} &  &|[name=fpe]| \text{KME$^2$} \\
    };
    \draw[->] 
    (nmqj) edge node[left] {$\mathcal{P}(\hi)$} (lme)
    (nmqsd) edge node[left] {$\mathcal{P}(\hi)$} (fpe)
    ;
    \draw (nmqj)  edge[->] node[above] {$\varepsilon\to 0$} (nmd) ;
    \draw (lme) edge[->] node[above] {$\varepsilon\to 0$} (fpe);
  \end{tikzpicture}
  \caption{Relation between non-Markovian quantum jumps (NMQJ),
    non-Markovian quantum diffusion (NMQD) and 
    non-Markovian quantum state diffusion (NMQSD).  
    In $\mathcal{P}(\hi)$, NMQJ corresponds to a Liouville master equation (LME), 
    whereas NMQSD is associated with  a 2nd order Kramers-Moyal expansion
    (KME$^2$)
    of the LME. In other words, the diffusive limit can be taken in such a 
    way that the LME associated with NMQJ 
    transforms to a KME$^2$  associated with NMQSD.
    However, when the very same limit is taken in $\hi$,
    it results to a completely new unraveling, which we call 
    non-Markovian Quantum Diffusion (NMQD).\label{fig:diagram}}
\end{figure}

On {the} theoretical side, our motivation is to look for the missing connection between the quantum jump \cite{PhysRevLett.100.180402} 
and quantum state diffusion \cite{PhysRevLett.82.1801} approaches
in the non-Markovian regime -- and with the help of these results expand significantly their applicability 
{of the former} for complex practical problems. First, we formulate  both approaches  in the projective Hilbert
space, thus extending the well known results from the Markov \cite{PhysRevLett.74.3788}
to the non-Markovian regime. Then a diffusive limit of the quantum jumps is taken in such a way
 that it coincides with quantum state diffusion in the projective Hilbert space and in 
the non-Markovian regime. Interestingly, the same limit in Hilbert space   
results in a completely new unraveling, which we call non-Markovian quantum 
diffusion (see Fig.~\ref{fig:diagram}). We enhance the quantum jumps and 
quantum diffusion approaches with kernel smoothing techniques~\cite{ghosh2017kernel},
which allows us to handle driven dissipative systems with quantum memory effects 
easily. Lastly, we apply all of the methods to the {driven dissipative 
two level atom}. 
%Our results pave the 
%way for solving important problems of current interest related to the latest 
%state-of-the-art experiments on complex open quantum systems.

\paragraph{Open systems and projective Hilbert space.---}
A typical model for open systems in the interaction picture
with respect to environment Hamiltonian $H_B=\sum_\lambda \omega_\lambda a_\lambda^\dagger a_\lambda$ is 
\begin{align}\label{eq:model}
  H(t)=& H_S(t)+\sum_\lambda g_\lambda L a_\lambda^\dagger e^{i\omega_\lambda t}+g_\lambda^*L^\dagger a_\lambda e^{-i\omega_\lambda t},
\end{align}
where the creation- and annihilation operators $a_\lambda^\dagger$ and $a_\lambda$ of 
a bath mode labeled by $\lambda$ satisfy the bosonic commutation relations 
$[a_\lambda,a_\mu^\dagger]=\delta_{\lambda\mu}$.
We assume  that the coupling operator $L$ is 
traceless, i.e. $\tr{L}{}=0$. 
%This type Hamiltonian can model many different  types 
%of open quantum systems, such as cold atomic gases, single atoms or trapped ions coupled to cavity 
%modes or a molecular aggregate, for example.
%According to an operational viewpoint to quantum mechanics, measurement outcome
%probabilities, produced by the famous Born rule 
%are the outputs 
%of the theory which can be compared with experimental data~\cite{Busch:2016:QM:3074076}.
 % Above, $\rho$ is 
% a positive trace class operator with unit trace  and $E_i$ a POVM element.
%However, any mixed state can be decomposed in arbitrarily many different ways 
%in terms of pure states.  The Born rule stays intact if we scale the pure state
%vectors in the decomposition 
%with arbitrary non-zero complex number and compensate the scaling in the 
%probability distribution over the pure states. 
%A description which is invariant with respect to this scaling 
%takes places 
In a projective Hilbert space $\mathcal{P}(\hi)$,
each point is associated with a projector $\kb{\psi}{\psi}$~\cite{breuer2002theory,chruscinski2004geometric}. 
Given a separable Hilbert space, 
coordinates $\psi_i\in\complex$ on a $\mathcal{P}(\hi)$ can be easily constructed 
with respect to a fixed orthonormal basis  as $\ket{\psi}=\sum_i\psi_i\ket{i}$. % For our purposes, it is important to show that 
% the Born rule remains intact under scaling of pure state vectors accounting for many different ways a mixed state can be written.
For more information on $\mathcal{P}(\hi)$, see the Supplementary Material~\cite{Note1}.
%Next we introduce 
%non-Markovian quantum state diffusion \cite{PhysRevLett.82.1801}and non-Markovian quantum jumps 
%methods \cite{PhysRevLett.100.180402}.

\paragraph{Non-Markovian quantum state diffusion.---}
Reduced system dynamics can be represented exactly 
for a large class of models, even   
beyond Eq.~(\ref{eq:model})~\cite{PhysRevLett.119.180401}, with the
following linear non-Markovian quantum state 
diffusion (NMQSD)  equation
\begin{align}\label{eq:nmqsd}
  \partial_t\ket{\psi(t,z^*)}=&\left[-i H_S(t) +z_t^*L\right]\ket{\psi(t,z^*)}\notag\\
  &-L^\dagger\int\limits_0^t\diff s\, \alpha(t-s)\frac{\delta\ket{\psi(t,z^*)}}{\delta z_s^*}.
\end{align}
Here, $L$ is the coupling operator between the system and the bath and 
$H_S(t)$ is an arbitrary Hamiltonian acting on the open system~\cite{PhysRevA.58.1699,PhysRevLett.82.1801}.
NMQSD is driven by a complex valued colored Gaussian noise $z_t^*$, completely
characterized by the correlations
\begin{align}\label{eq:correlations}
  \mean{z_tz_s^*}=\alpha(t-s),\qquad\mean{z_t^*}=\mean{z_tz_s}=0,
\end{align}
where 
$\mean{\cdot}$ is the average over the noise process $z_t^*$. Solutions 
$\ket{\psi(t,z^*)}$ 
are analytic functionals of the whole noise process $z_t^*$ up 
to time $t$.  

In the remainder of this Letter, we will make the following 
restriction. 
We assume that the functional 
derivative satisfies, at least 
approximately~\cite{PhysRevA.60.91}
\begin{align}
\label{eq:restriction}
  \frac{\delta}{\delta z_s^*}\ket{\psi(z^*,t)}=f(t,s)L\ket{\psi(z^*,t)}.
\end{align}
Eq.~(\ref{eq:restriction}) guarantees
that the mean state will evolve according to a closed form master equation. However, the 
NMQSD method itself works perfectly well even if no such master equation
exist for the reduced state.
% Then the above equation simplifies to 
% \begin{align}\label{eq:nmqsd_simple}
%   \partial_t\ket{\psi(z^*,t)}
%   &=\left[-i H_S(t) +z_t^*L-F(t)L^\dagger L\right]\ket{\psi(z^*,t)}
% \end{align}
%with $F(t)=\int_0^t\diff s\, \alpha(t-s)f(t,s)$.

The above stochastic Schrödinger equation~(\ref{eq:nmqsd}) 
satisfies the ordinary rules of calculus since the noise process has a finite correlation time.  
%The white noise limit of Eq.~(\ref{eq:nmqsd})  
%needs to interpreted in the Stratonovich sense.
% Since Eq.~(\ref{eq:nmqsd}) is a linear equation, the solution is 
% given by a time ordered exponential
% \begin{align}
%   \ket{\psi(z^*,t)} = T\exp\int\limits_0^t\diff s\,\left[-i H_S(t) +z_t^*L-F(t)L^\dagger L\right]\ket{\psi(0)}.
% \end{align}
The dynamics of the average state $\rho(t)=\mean{\kb{\psi(z^*,t)}{\psi(z^*,t)}}$
described by Eq.~(\ref{eq:nmqsd}) with assumption 
(\ref{eq:restriction}) reads
\begin{align}\label{eq:master_eq}
  \dot{\rho}(t)&= -i[H_S(t)+S(t)L^\dagger L,\rho(t)]
                  +2\gamma(t)L\rho(t) L^\dagger \notag\\
  &- \gamma(t)\left\{L^\dagger L,\rho(t)\right\},
\end{align}
where $F(t)=\gamma(t)+iS(t)$ and $F(t)=\int_0^t\diff s\, \alpha(t-s)f(t,s)$.
%Quantum memory effects are described by the colored noise $z_t^*$ and the 
%functional derivate term under the memory integral 
%in Eq.~(\ref{eq:nmqsd}), which describes how a particular  trajectory depends 
%on other stochastic trajectories nearby.

To look for a connection between NMQSD and non-Markovian
quantum jumps, we first have to derive a representation of the former
in the projective Hilbert space. The probability
density functional can be expressed as  
\begin{align}\label{eq:P_Q}
  P_Q[\psi,t] & = \mean{\delta(\psi-\psi(z^*,t))}.
\end{align}
We show in Sec.~\ref{sec:deriv-fokk-planck} of~\footnote{Supplementary Material}, that the probability density functional 
satisfies the following second order partial differential equation 
\begin{align}\label{eq:LNMQSD_Fokker_Planck}
  \partial_t P_Q[\psi,t] =&\sum_{k=1}^d \partial_{\psi_k}c_k(\psi)P_Q[\psi,t]
    +\partial_{\psi_k^*}c_k^*(\psi)P_Q[\psi,t]\notag\\
  &+\sum_{k,l=1}^d \partial^2_{\psi_k\psi_l^*}d_{kl}(\psi)P_Q[\psi,t],
  % =-\sum_{k=1}^d\partial_{\psi_k}\bra{k}\left(-iH-F(t)L^\dagger L\right)\ket{\psi}P_Q[\psi,t]
  %  -\sum_{k=1}^d\partial_{\psi_k^*}\bra{\psi}\left(iH-F^*(t)L^\dagger L\right)\ket{k}P_Q[\psi,t]\notag\\
  %  &+\sum_{k,l=1}^d\partial_{\psi_k\psi_l^*}^2\left(F(t)+F^*(t)\right)\bra{k}L\ket{\psi}\bra{\psi}L^\dagger\ket{l}P_Q[\psi,t],
\end{align}
where the drift and diffusion coefficients are 
$c_k(\psi)=\bra{k}\left(-iH-F(t)L^\dagger L\right)\ket{\psi}$ and 
$d_{kl}(\psi)=\left(F(t)+F^*(t)\right)\bra{k}L\ket{\psi}\bra{\psi}L^\dagger\ket{l}$, respectively.
{NMQSD thus corresponds to a $2$nd order Kramers - Moyal expansion 
in $\mathcal{P}(\hi)$~\cite{risken2012fokker}. 
If the diffusion coefficient $F(t)+F^*(t)=2\gamma(t)$ is not negative for any 
time $t$ the KME$^2$ equation is, in fact, a proper Fokker-Planck equation~\cite{gardiner2009stochastic}.}

\paragraph{Non-Markovian Quantum Jumps.---}
Master equations of the form 
\begin{align}\label{eq:master_eq2}
  \dot{\rho}(t)=& -i[H_S(t)+\sum_ks_k(t)L_k^\dagger L_k,\rho]
                  +\sum_k2\gamma_k(t)L_k\rho L_k^\dagger \notag\\
                  &- \gamma_k(t)\left\{L_k^\dagger L_k,\rho(t)\right\},
\end{align}
can be unravelled with non-Markovian quantum jumps (NMQJ)~\cite{PhysRevLett.100.180402,PhysRevA.79.062112,H_rk_nen_2010},
\footnote{We demand that at time $t=0$ all of the decay rates $\gamma_k(t)$ have to be non-negative}. 
It is a piecewise deterministic process in the Hilbert
space of the open system. Here we present a linear
version of the process  (LNMQJ) given by the following Ito stochastic differential equation
\begin{align}\label{eq:nmqj_linear}
  \ket{\diff\psi}
  =&-i G(t)\ket{\psi}\diff t+
  \sum_k(L_k-\id)\ket{\psi}\diff N_+^k(t)\notag\\
     &+\int\diff\psi'\,\left(\ket{\psi'}-\ket{\psi}\right)\diff N_{-,\psi'}^k(t),
\end{align}
where $G(t)= H_S(t)+\sum_k s_k(t)L_k^\dagger L_k-i \gamma_k(t)[L_k^\dagger L_k-\id]$.
Increments of the Poisson processes, $\diff N_+^k(t)$ and $\diff N_{-,\psi'}^l(t),$
are mutually independent 
  $\diff N_+^k(t)\diff N_+^l(t)=\delta_{kl}\diff N_+^k(t)$,
  $\diff N_{-,\psi'}^k(t)\diff N_{-,\psi''}^l(t)=\delta_{kl}\delta(\psi'-\psi'')\diff N_{-,\psi'}(t)$
  and $\diff N_+^k(t)\diff N_{-,\psi'}^l(t)=0$.
The mean values of the increments are
  $\Ex{\diff N_+^k(t)}=2\gamma_+^k(t)\diff t$ and 
  $\Ex{\diff N_{-,\psi'}^l(t)}=2\gamma_-^l(t)\frac{P[\psi',t]}{P[\psi,t]}\delta(\psi-L_l\psi')\diff t$,
where $\gamma_k(t)=\gamma_k^+(t)-\gamma_k^-(t)$. It is easy to see that 
the average evolution reproduces Eq.~(\ref{eq:master_eq2}). 

In NMQJ, the 
memory effects reside in the jump probability from a source state 
$\psi$ to a target state $\psi'$ via channel $k$ when $\gamma_k(t)<0$. In 
particular, a ``reverse jump'' can occur from $\psi$ to $\psi'$ iff  
$L_k\ket{\psi'} = \ket{\psi}$. The probability of such jumps depends on the 
ratio $P[\psi',t]/P[\psi,t]$. 
%Thus the reverse jump updates the ensemble in such 
%a way that it increases the probability to be in a state which could decay 
%via channel $k$ when $\gamma_k(t)>0$. 
In order to compute the jump probability,
the knowledge of the whole ensemble is required~\cite{PhysRevLett.100.180402}. 
This poses a serious challenge  
since a state $\kb{\psi}{\psi}$ has measure zero in $\mathcal{P}(\hi)$. We  describe later a 
method to overcome this.

Now, Eq.~(\ref{eq:master_eq}) is 
equivalent to Eq.~(\ref{eq:master_eq2}) with 
$2m\, (1\leq k \leq 2m)$ time dependent rates and time independent jump 
operators defined as
\begin{align}\label{eq:scaling}
  s_k(t)=&\frac{s(t)}{2m\abs{\xi_k}^2\varepsilon^2},\qquad
  \gamma_k(t)=\frac{\gamma(t)}{2m\abs{\xi_k}^2\varepsilon^2},\notag\\
  L_k=&\id+\varepsilon\xi_kL,\,\,{\rm s.t.}\, \xi_k+\xi_{k+m}=0,
\end{align}
where $\xi_k\in\complex$, $\abs{\xi_k}=\abs{\xi}$ and $\varepsilon>0$.
The deterministic part $G(t)$ of the  quantum jump process in Eq.~(\ref{eq:nmqj_linear})
transforms under (\ref{eq:scaling}) to $G'(t)=H_S(t)+s(t)L^\dagger L-i\gamma(t)L^\dagger L+\Theta(t)\id$.
%\begin{align}
 % G'(t)=H_S(t)+s(t)L^\dagger L-i\gamma(t)L^\dagger L+\Theta(t)\id.
%\end{align}
The last term $\Theta(t)=\sum_{k=1}^{2m}\frac{s(t)}{2m\abs{\xi_k}^2\varepsilon^2}$ 
is a global phase factor, which can be neglected.
If $\norm{\varepsilon L}< 1$, then operators $L_l$ are invertible
\footnote{${L_l}^{-1}=\sum_{j=0}^\infty (-1\xi_l\varepsilon)^jL^j=\id-\varepsilon\xi_lL+\mathcal{O}(\varepsilon^2)$, 
when ${\norm{\varepsilon L}} < 1$.}.
%By assuming that we have chosen $\varepsilon$ such that 
%the invertibility is guaranteed
% we can % use the rule $\delta(\vec f(\vec x))=
% \sum_{\vec x_0|\vec f(\vec x_0)=0}\frac{\delta(\vec x-\vec x_0)}{\abs{J(\vec x_0)}}$ for 
% multidimensional Dirac $\delta$-functions, where $J_{ij}=\frac{\partial f_i(\vec x)}{\partial x_j}$ (Jacobian), to
%write $\delta(\psi-L_l\psi')=\delta(L_l^{-1}\psi-\psi')(\abs{\det L_l}\abs{\det L_l^\dagger})^{-1}$. 
In this case, the transformed statistics of the Poisson increments eventually become  
\begin{align}\label{eq:scaled_poisson_statistics}
  \Ex{\diff N_+^k(t)}&=\frac{\gamma_+(t)}{m\varepsilon^2\abs{\xi_k}^2}\diff t,\notag\\
 % \Ex{\diff N_{-,\psi'}^l(t)}&=\frac{\gamma_-(t)}{m\varepsilon^2\abs{\xi_k}^2}
  %                           \frac{P[\psi',t]}{P[\psi,t]}
  %                           \delta(\psi-L_l\psi')\diff t.,\\
  \Ex{\diff N_{-,\psi'}^l(t)}&=\frac{\gamma_-(t)}{m\varepsilon^2\abs{\xi_k}^2}
                               \frac{P[\psi',t]}{P[\psi,t]}
                               \frac{\delta(L_l^{-1}\psi-\psi')}{\abs{\det L_l}\abs{\det L_l^\dagger}}\diff t.
\end{align}
%where the both $\det L_l$ and $\det L_l^\dagger$ appear 
%due to the definition of the $\delta$-function (see Sec.~\ref{sec:dirac-delta-function} in~\cite{Note1}).
% Therefore, we can write Eq.~(\ref{eq:scaled_poisson_statistics}) as 
% \begin{align}
 % 
% \end{align}
Remarkably, after the transformation the increment 
$\diff N_{-,\psi'}^l(t)$ does not 
depend on the target state of the jump,$\ket{\psi'}$, anymore. This arises because a
reverse jump corresponds to a mapping 
$\ket{\psi}\mapsto L_l^{-1}\ket{\psi}=\ket{\psi'}$, i.e.
the target state of the 
jump is given by the action of the inverse operator on the 
source state $\ket{\psi}$. 

Therefore, we can write the transformed process as
%The transformed process thus can be written as 
\begin{align}\label{eq:lnmqj_transformed}
  \ket{\diff\psi_t}
  =&-i G'(t)\ket{\psi}\diff t+
  \sum_k\Bigg[(L_k-\id)\ket{\psi_t}\diff M_+^k(t)\notag\\
     &+\left({L_k}^{-1}-\id\right)\ket{\psi_t}\diff M_-^k(t)\Bigg],
\end{align}
with mutually independent Poisson increments $\diff M_\pm^k$ with 
statistics $\Ex{\diff M_+^k(t)}=\frac{\gamma_+(t)}{m\varepsilon^2\abs{\xi_k}^2}\diff t$ and 
 $\Ex{\diff M_{-}^k(t)}=\frac{P[{L_k}^{-1}\psi,t]}{P[\psi,t]\abs{\det L_k}\abs{\det L_k^\dagger}}\frac{\gamma_-(t)}{m\varepsilon^2\abs{\xi_k}^2}\diff t$,
which are just relabeled increments of Eq.~(\ref{eq:scaled_poisson_statistics}).
To assert that this equation is still valid, we compute the 
average evolution of $\kb{\psi}{\psi}$ which coincides 
with the master equation (\ref{eq:master_eq}) (see  Sec.~\ref{sec:average-evolution} of the~\cite{Note1}).

It is worth stressing that when $\gamma_k(t)<0$, the 
quantum jumps are given by the inverse jump operator $L_k^{-1}$. 
Contrary to the original approach in~\cite{PhysRevLett.100.180402}, the
quantum jumps and reverse quantum jumps are exactly inverses of each other.
%This means that the reverse jump occuring when $\gamma_k(t)<0$ is
%exactly the reverse operation of a jump occuring when $\gamma_k(t)>0$.
The quantum memory effects are contained in the probability for these jumps which 
still depends on the ratio $P[L_k^{-1}\psi,t]/P[\psi,t]$.
\paragraph{LNMQJ in projective Hilbert space.---}
In the projective Hilbert space LNMQJ corresponds to the 
following Liouville master equation~\cite{H_rk_nen_2010}
\begin{align}\label{eq:LNMQJ_Liouville}
  \partial_t P[\psi,t] 
  &= i\sum_k\partial_{\psi_k}\left(\bra{k}G'(t)\ket{\psi}P[\psi,t]\right)\notag\\
  &-\partial_{\psi_k^*}\left(\bra{\psi}G'^\dagger(t)\ket{k}P[\psi,t]\right)\notag\\
  &+\int\diff\phi\, \left(R[\psi|\phi]P[\phi,t]-R[\phi|\psi]P[\psi,t]\right),
\end{align}
where the jump rates $R[\phi|\psi]$ are
\begin{align}
  R[\phi|\psi] =& \sum_{k=1}^{2m} \frac{\gamma_+(t)}{m\varepsilon^2\abs{\xi_k}^2}\delta(\phi-L_k\psi)\notag\\
  &+\frac{\gamma_-(t)}{m\varepsilon^2\abs{\varepsilon_k}^2}\frac{P[\phi,t]}{P[\psi,t]}\delta(\psi-L_k\phi).
\end{align}
When comparing the drift terms in Fokker-Planck equation~(\ref{eq:LNMQSD_Fokker_Planck}) and
in the Liouville master equation~(\ref{eq:LNMQJ_Liouville}), we see that they are equal.
The jump part takes the form
$\int\diff\phi\, \left(R[\psi|\phi]P[\phi,t]-R[\phi|\psi]P[\psi,t]\right)
=\sum_{k=1}^{2m}\frac{\gamma(t)}{m\varepsilon^2\abs{\xi_k}^2}F_k[\psi]
-\frac{2\gamma(t)}{\varepsilon^2\abs{\xi}^2}P[\psi,t]$, 
where
\begin{align}
F_k[\psi]=
\frac{P[L_k^{-1}\psi,t]}{\abs{\det L_k}\abs{\det L_k^\dagger}}.
\end{align}
 After expanding 
$F_k[\psi]$ to second order in $\varepsilon$ and 
assuming $m>2$ we find 
%\begin{align}
 $ \int\diff\phi\, \left(R[\psi|\phi]P[\phi,t]-R[\phi|\psi]P[\psi,t]\right) 
  {\to}$\\$ \sum_{k,l=0}^d \partial^2_{\psi_k\psi_l^*}\left(2\gamma(t)\bra{k}L\ket{\psi}\bra{\psi}L^\dagger\ket{l}P[\psi,t]\right)$,
while $\epsilon\to 0$
\footnote{For the computation we use 
the results of  ~\cite{Note1} to compute the determinant,
to expand the probability density, to expand a rational polynomial
and  choose $\xi_k=e^{i\frac{\pi}{m}(k-1)}$ with $m\geq 2$}.
%\end{align}
We thus have proven the validity of the part 
$\mathrm{LME}\xrightarrow{\varepsilon\to 0}\mathrm{FPE}$ of the diagram in Fig.~\ref{fig:diagram}.

\paragraph{Non-Markovian quantum diffusion.---}
Next we take the above diffusion limit directly on 
the piecewise deterministic LNMQJ process in the Hilbert space. Full details can be found in
the Supplement~\cite{Note1}.
First, Eq.~(\ref{eq:lnmqj_transformed}) is expanded to first order 
in $\varepsilon$, resulting in  % \footnote{We use  $L_k=\id+\varepsilon\xi_k L+\mathcal{O}(\varepsilon^2)$ and
 % $(L_l^{-1}-\id)=-\varepsilon\xi_l L+\mathcal{O}(\varepsilon^2)$}
\begin{align}\label{eq:lnmqj_appprox}
  \ket{\diff\psi_t}
  =&-i G'(t)\ket{\psi}\diff t+
  \sum_k\Bigg[\xi_kL\ket{\psi_t}\varepsilon\diff M_+^k(t)\notag\\
  &-\xi_kL\ket{\psi_t}\varepsilon\diff M_-^k(t)\Bigg]+\mathcal{O}(\varepsilon^2).
\end{align}
We define  new processes
$\diff  V_{\pm}^k=\varepsilon\diff M_{\pm}^k-\varepsilon E\left[\diff M_\pm^k\right]$ \cite{doi:10.1063/1.3263941}
and by using the Ito rules, we have $\Ex{\diff V_{\pm}^k}=0$, $\Ex{\diff V_{-}^k\diff V_{+}^l}=0$ and 
 $\Ex{\diff V_{+}^k\diff V_{+}^l}=\delta_{kl}(\varepsilon \diff V_{+}^k+\frac{\gamma_+(t)}{m\abs{\xi_k}^2}\diff t)$.
 We then define 
$\lim_{\varepsilon\to 0}\diff V_{\pm}^k=\diff W_\pm^k$,
where the increments $\diff W_\pm^k$ satisfy the following Ito rules 
\begin{align}
  &\Ex{\diff W_{\pm}^k}=0,\qquad \Ex{\diff W_{-}^k\diff W_{+}^l}=0,\notag\\
  &\Ex{\diff W_{+}^k\diff W_{+}^l}=\delta_{kl}\frac{\gamma_+(t)}{m\abs{\xi_k}^2}\diff t,\\
  &\Ex{\diff W_{-}^k\diff W_{-}^l}
  =\delta_{kl}\frac{\gamma_-(t)}{m\abs{\xi_k}^2}\diff t.\notag
\end{align}
The goal is now to express the stochastic Schrödinger equation~(\ref{eq:lnmqj_appprox})
in terms of Wiener increments $\diff W_\pm^k$. After some simplification steps (detailed in the Supplement~\cite{Note1}), we find 
% The jump part of the above stochastic equation (\ref{eq:lnmqj_appprox}) 
% in terms of $\diff V_\pm^k$ can be written as
% $\sum_k\big[\xi_kL\ket{\psi}\varepsilon\diff M_{+}^k(t)-\xi_kL\ket{\psi}\varepsilon\diff M_{-,\varepsilon}^k(t)\big]
% =\sum_k\bigg[\xi_k\left(\diff V_{+}^k-\diff V_{-}^k\right)
% +\frac{\gamma_-(t)\xi_k}{m\abs{\xi_k}^2P[\psi,t]}\left(A_k[\psi]+\mathcal{O}(\varepsilon)\right)\diff t\bigg]L\ket{\psi}$,
% where the last equality follows from $\xi_k+\xi_{k+m}=0$ and $\abs{\xi_k}=\abs{\xi}$, $\forall k$. 
% % which destroys the term $\sum_{k=1}^{2m}\Ex{\diff M_k^+(t)}$ and 
% % zeroth order term in $\varepsilon$ from $\sum_{k=1}^{2m}\Ex{\diff M_k^-(t)}$.
% We can further simplify the expression by noting
% %\begin{align}
%   $\sum_k\frac{\gamma_-(t)\xi_k}{m\abs{\xi_k}^2P[\psi,t]}A_k[\psi]
%   =\sum_k\frac{\gamma_-(t)\xi_k}{m\abs{\xi_k}^2P[\psi,t]}
%   \sum_{n=0}^d\xi_k\bra{n}L\ket{\psi}\frac{\partial P[\psi,t]}{\partial \psi_n}+\xi_k^*\bra{\psi}L^\dagger\ket{n}\frac{\partial P[\psi,t]}{\partial\psi_n^*}
%   =2\frac{\gamma_-(t)}{P[\psi,t]}
%   \sum_{n=0}^d\bra{\psi}L^\dagger\ket{n}\frac{\partial P[\psi,t]}{\partial\psi_n^*}=2{\gamma_-(t)}
%   \sum_{n=0}^d\bra{\psi}L^\dagger\ket{n}\frac{\partial \ln P[\psi,t]}{\partial\psi_n^*}$,
% %\end{align}
% if we assume $\sum_k \xi_k^2 = 0$ \footnote{An example choice for $\xi_k$'s is  
% $\xi_1 = 1$, $\xi_2 = -1$, $\xi_3 = i$ and $\xi_4 = -i$}.
in the limit $\varepsilon\to 0$ %we can write (with the above choice for $\xi_k$) 
\begin{align}\label{eq:NMQD}
  \ket{\diff \psi} =&\bigg(-iG'(t)+2\gamma_-(t)\sum_{n=0}^d\bra{\psi}L^\dagger\ket{n}\frac{\partial \ln P[\psi,t]}{\partial\psi_n^*}L\bigg)\ket{\psi}\diff t\notag\\
  &+L\ket{\psi}\diff Z_+ -L\ket{\psi}\diff Z_-,
\end{align}
where $\diff Z_\pm = \sum_k\xi_k\diff W_{\pm}^k$. The complex noise $\diff Z_\pm$ satisfies 
\begin{align}
  &\Ex{\diff Z_\pm}=0,\quad \Ex{\diff Z_\pm\diff Z_\pm} = 0,\notag\\
  &\Ex{\diff Z_\pm\diff Z_\pm^*} = 2\gamma_\pm(t)\diff t.
\end{align}
The average evolution of NMQD equation (\ref{eq:NMQD}) corresponds to 
Eq.~(\ref{eq:master_eq2}) as we show in the Supplement~\cite{Note1}.

Interestingly, both noises $\diff Z_\pm$ couple to the system via $L$ but with 
a different phase. Nevertheless, both noise terms produce  ``sandwich''  terms
$2\gamma_{\pm}(t)L\rho L^\dagger\diff t$ on the average evolution.  The drift term with logarithmic derivative 
compensates the term $2\gamma_{-}(t)L\rho L^\dagger\diff t$ on average such that 
the correct sandwich term $2(\gamma_+(t)-\gamma_-(t))L\rho L^\dagger\diff t$ emerges.
The term proportional to the logarithmic derivative can be 
seen as the change in the stochastic entropy of the system which contributes to  
the deterministic evolution \cite{PhysRevLett.95.040602}. 

\paragraph{Kernel smoothing.---}
A Gaussian kernel $K$ is defined 
\begin{align}\label{eq:Gaussian-kernel}
  K[\psi] = \frac{1}{\pi^{d+1}}e^{-\norm{\psi}^2}, \ket{\psi}\in\complex^{d+1}.
\end{align}
Given an ensemble of stochastic states $\{\ket{\psi^{\nu}}\}_{\nu=1}^M$, we 
estimate the probability density $P[\psi]$ in the projective 
Hilbert space with 
\begin{align}\label{eq:kernel_estimate}
  P_\sigma[\psi] = \frac{1}{M(\sigma^2\pi^{d+1})}\sum_{\nu=1}^MK[(\psi-\psi^\nu)/\sigma],
\end{align}
where $\sigma>0$ is a free parameter. A rule of thumb for choosing the variance 
is  that $\sigma = M^{\frac{-1}{d+5}}$~\cite{ghosh2017kernel}, where $d$ is the 
real dimension of the underlying Hilbert space. Using the estimated density, we can compute
the logarithmic derivative of the density appearing in Eq.~(\ref{eq:NMQD}) as 
\begin{align}
  \frac{\partial \ln P_\sigma[\psi]}{\partial \psi_n^*}
  =-\frac{\sum_{\nu=1}^Me^{-\norm{\psi_n-\psi_n^\nu}^2}\frac{\psi_n-\psi_n^\nu}{\sigma^2}}{\sum_{\nu=1}^Me^{-\norm{\psi_n-\psi_n^\nu}^2}}
  =-\frac{\psi_n -\langle\langle \psi_n\rangle\rangle}{\sigma^2},
\end{align}
where average $\langle\langle\cdot\rangle\rangle$ is taken with respect 
to distribution  $p_\sigma = \frac{1}{\mathcal{Z}}e^{-\frac{\norm{\psi-\psi^\nu}^2}{\sigma^2}}$,
with $\mathcal{Z} = \sum_{\nu=1}^Me^{-\norm{\psi_n-\psi_n^\nu}^2}$.
Kernel estimation can be also used to evaluate the ratios 
$\frac{P_\sigma[\psi']}{P_\sigma[\psi]} = \frac{\sum_{\nu'=1}^M K[(\psi'-\psi^{\nu'})/\sigma]}{\sum_{\nu=1}^M K[(\psi-\psi^{\nu})/\sigma]}$.
Therefore, after performing the transformation (\ref{eq:scaling}) on the NMQJ and 
using the smoothed estimate for $P[\psi,t]$ we can compute the reverse jump probabilities easily.
The reason for this simplification is that the target state of the jump is directly given by the inverse jump operator and 
the ratio of probabilities for the target and the source state to occur can be efficiently evaluated 
from the estimate.

\paragraph{{Example: Driven dissipative two level atom.}---}
An open system with $H_S=\frac{\omega}{2}\sigma_z + \frac{\Omega}{2}\sigma_x$ and 
$L=\sigma_-$ corresponds to {an amplitude damped two level atom 
with driving and is not solvable in closed form.} 
We assume that the bath correlation function takes the 
following exponential form
\begin{align}
  \alpha(t,s)=g\frac{\Gamma}{2}e^{-i\omega_c(t-s)-\Gamma\abs{t-s}},
\end{align}
where $\Gamma$ is the inverse of the bath correlation time $\tau_c=\Gamma^{-1}$, 
$\omega_c$ is the bath resonance frequency and $g>0$ is a dimensionless parameter describing the 
overall system bath coupling strength. The limit $\Gamma\to\infty$ leads to a singular 
bath correlation function $\alpha(t,s)\to g\delta(t-s)$ and to a 
Gorini-Kossakowski-Sudarshan-Lindblad master equation with time independent decay rate 
$g$~\cite{Gorini1976,Lindblad1976}.  The chosen correlation function  can emerge from a microscopic model 
where the driven two level system is placed inside a 
leaky cavity near absolute zero temperature such that thermal excitations can 
be neglected.
When the bath correlation time is short, 
Eq.~(\ref{eq:restriction}) is approximately true~\cite{PhysRevA.60.91}. 
% we can replace in the non-Markovian QSD
%approach the functional derivative with~\cite{PhysRevA.60.91}
%\begin{align}
  %&\int_0^t\diff s\, \alpha(t-s)\frac{\delta}{\delta z_s^*}\ket{\psi_t(z^*)}
 % \approx \int_0^t\diff s\, \alpha(t-s)\sigma_-\ket{\psi_t(z^*)}\notag\\
 % &=F(t)\sigma_-\ket{\psi_t(z^*)}.
%\end{align}
Within this approximation the NMQSD equation takes the following form 
\begin{align}
  \partial_t\ket{\psi_t(z^*)} =& -i H_S\ket{\psi_t(z^*)}+z_t^*\sigma_-\ket{\psi_t(z^*)}\notag\\
  &-F(t)\sigma_+\sigma_-\ket{\psi_t(z^*)},
\end{align}
with $\alpha(t,s)=\langle z_tz_s^*\rangle$ being the only non-zero correlation of the complex noise.
Then the average state obeys the following master equation
\begin{align}\label{eq:driven_tla_me}
  \partial_t \rho =& -i[\frac{\omega}{2}\sigma_z + \frac{\Omega}{2}\sigma_x+s(t)\sigma_+\sigma_,\rho]
                     +2\gamma(t)\sigma_+\rho\sigma_-\notag\\
  &-\gamma(t)\{\sigma_+\sigma_-,\rho\},
\end{align}
where $\gamma(t)+is(t)=F(t)$. 
% \begin{figure}[ht]
%   \includegraphics[width=0.49\textwidth]{decayrate}
%   \caption{\label{fig:decayrate} Decayrate $\gamma(t)=\frac{1}{2}(F(t)+F^*(t))$ for 
%     $\omega/\Gamma = 2$, $\omega_c/\Gamma = 5.5$, $\Omega/\Gamma=2$, $g=8$ plotted 
%     as a function of dimensionless time $\Gamma t$. When $\frac{1}{2} < \Gamma t< 3/2 $ 
%     decayrate is temporarily negative.}
% \end{figure}
\begin{figure}[t]
  \includegraphics[width=\linewidth]{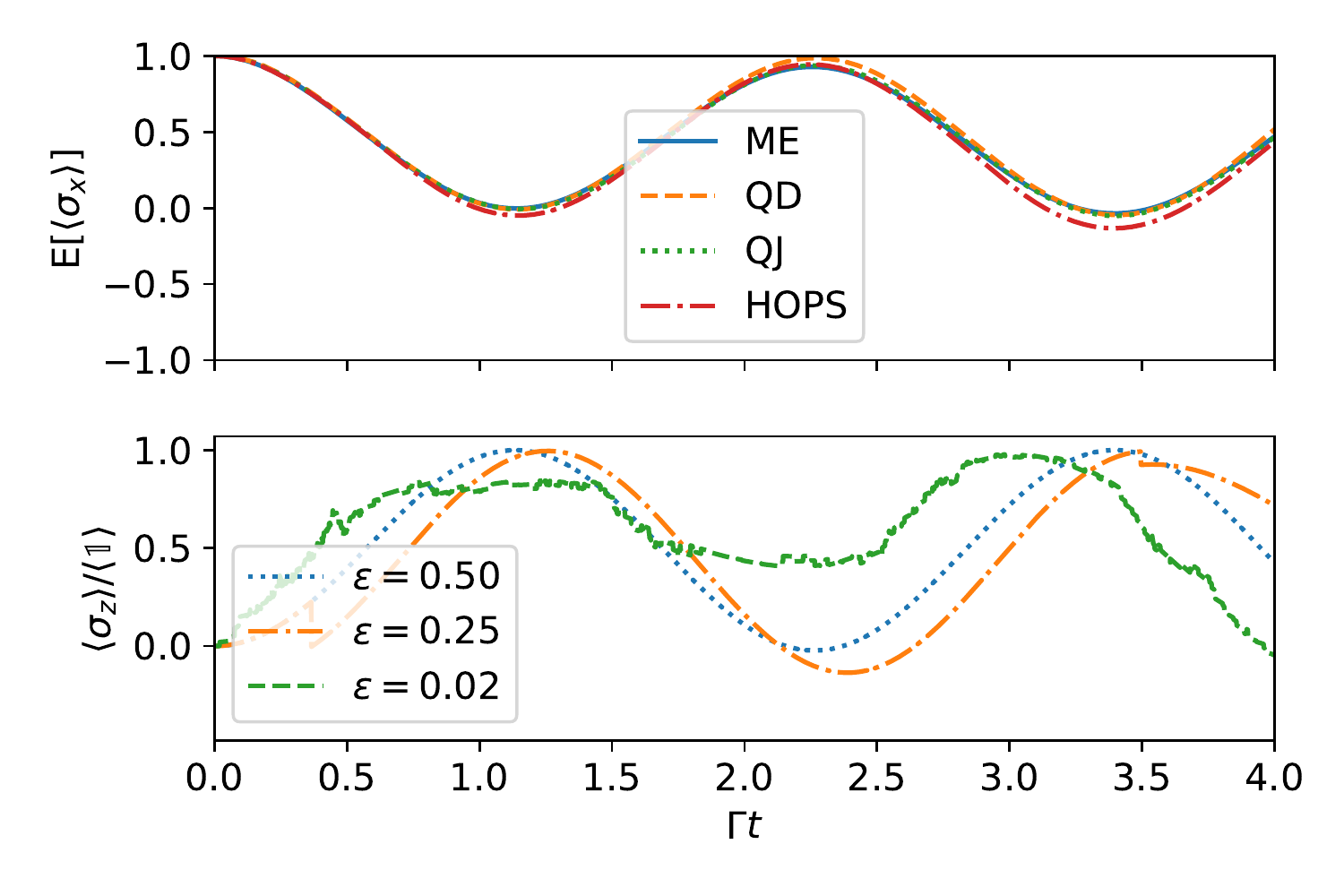}
  \caption{{\bf Top}: Ensemble average over 3000 stochastic trajectories 
  of $\langle \sigma_x\rangle$  computed with LNMQJ (dotted),  NMQD (dashed) with 
  $\varepsilon=\frac{1}{2}$ and HOPS (dash dotted) with comparison to the master 
  equation solution (ME). {HOPS is a numerically exact method and the 
  reasonable agreement shows that 
  the approximations leading to the master equation being unraveled are
  fairly consistent with the chosen parameters.} {\bf Bottom}: Normalized expectation value for $\sigma_z$ along a single 
  stochastic trajectory for different values of $\varepsilon$ using LNMQJ. The initial state is 
  $\ket{+}=\sqrt{\frac{1}{2}}(\ket{0}+\ket{1})$. \label{fig:example}
  %Other parameters are $\omega/\Gamma = 2$, 
  %$\omega_c/\Gamma = 5.5$, $\Omega/\Gamma=2$, $g=8$.
  }
  \label{fig:driven_tla}
\end{figure}
% \begin{figure}
%   \begin{minipage}{0.49\linewidth}
%     \includegraphics[width = \textwidth]{qsd}
%   \end{minipage}%
%   \begin{minipage}{0.49\linewidth}
%     \includegraphics[width = \textwidth]{lnmqj}
%   \end{minipage}\hfill
%   \begin{minipage}{0.49\linewidth} 
%     \includegraphics[width=\textwidth]{diffusive_limit}
%   \end{minipage}%
%   \begin{minipage}{0.49\linewidth}
%     \includegraphics[width=\textwidth]{vareps}
%   \end{minipage}
%   \caption{\label{fig:driven_tla} Ensemble averages of the Bloch vector components 
%   using NMQSD (top left), LNMQJ with $\varepsilon=\frac{1}{2}$ (top right) and NMQD (bottom left). 
%   In bottom right, we plot the $z$-component of the Bloch vector using LNMQJ 
%   for different values of $\varepsilon$.  The initial state is 
%   $\ket{+}=\sqrt{\frac{1}{2}}(\ket{0}+\ket{1})$. Other parameters are $\omega/\Gamma = 2$, 
%   $\omega_c/\Gamma = 5.5$, $\Omega/\Gamma=2$, $g=8$. In top row and bottom left we have 
%   averaged over $1000$ realizations.}
%\end{figure}
The LNMQJ unraveling (\ref{eq:driven_tla_me}), in turn, is  
\begin{align}
  \diff\ket{\psi}=& -i G(t)\ket{\psi}\diff t
  +\sum_{k=1}^4\varepsilon\xi_k\sigma_-\ket{\psi}\diff M_+^k(t)\notag\\ 
  &-\sum_{k=1}^4\varepsilon\xi_k\sigma_-\ket{\psi}\diff M_-^k(t),
\end{align}
where $\xi_1=1$, $\xi_2= -1$, $\xi_3=i$, $\xi_4=-i$ and 
$G(t)=(H_S-iF(t))\sigma_+\sigma_-$. The statistics of the Poisson increments are  
$\Ex{\diff M_+^k}=\frac{\gamma_+(t)}{2\varepsilon^2}\diff t$ and 
$\Ex{\diff M_-^k}=\frac{P[(\id-\varepsilon\xi_k\sigma_-)\psi,t]}{P[\psi,t]}\frac{\gamma_-(t)}{2\varepsilon^2}\diff t$.
Subsequently, the diffusive limit of LNMQJ process corresponding to the NMQD process for this system can be written as 
\begin{align}
  \diff\ket{\psi} =& \bigg(-iG(t)+2\gamma_-(t)\bra{\psi}\sigma_+\ket{0}\frac{\partial \ln P[\psi,t]}{\partial \psi_0^*}\sigma_-\bigg)\ket{\psi}\diff t\notag\\
  &+\sigma_-\ket{\psi}(\diff Z_+-\diff Z_-),
\end{align}
where zero mean complex noises satisfy the Ito rules  $\diff Z_\pm\diff Z_\pm^* = \gamma_\pm(t)\diff t$ and 
$\diff Z_\pm\diff Z_\pm=\diff Z_\mp\diff Z_\pm^*=0$.
We consider the following parameters in all of the numerical examples
$\omega/\Gamma = 2$, $\omega_c/\Gamma = 5.5$, $\Omega/\Gamma=0.5$, $g=0.8$ and 
we plot all dynamical quantities as a function of the dimensionless time $\Gamma t$. 
The decay rate $\gamma(t)$ is temporarily negative when $\frac{1}{2} < \Gamma t< 3/2 $ 
for these parameter values. {Figure~\ref{fig:example}} shows 
a good agreement between the master equation solution and its unravelings. However, 
we also solved the dynamics exactly using the HOPS approach to 
NMQSD~\cite{PhysRevLett.113.150403,Hartmann2017}. The small
disagreement
shows that the approximations leading to the master equation (\ref{eq:driven_tla_me}) 
are not fully consistent with the chosen parameters. Therefore, a word of caution is in place
here; within the master equation approach, the quality of the obtained 
equation is extremely hard to assess\footnote{{HOPS with hierarchy order 1 corresponds 
closely to the level of approximation we make when using- and unraveling the master
quation with respect to exact dynamics. In Fig.~\ref{fig:driven_tla} we have 
truncated the hierarchy after 8 levels.}}.
{In the bottom panel, we also show examples of
single trajectories with LNMQJ for different values of $\epsilon$.}
%Note, that the parameters we have chosen are not necessarily in the regime of validity 
%of the approximation performed to obtain the master equation (\ref{eq:driven_tla_me}).
%and we can not expect that our approximate models would correspond to the exact 
%microscopical dynamics. However, 
{The purpose is to demonstrate 
the agreement  of the ensemble averages with the master equation solution 
-- and that with our new methodological results,  even driven systems can be 
very easily handled with the jump method whenever a reliable master equation
is available.}
\paragraph{Conclusions.---}
We have provided a connection between quantum state diffusion and 
quantum jumps in the non-Markovian regime. As a by product of these investigations we 
introduced a linear version of the non-Markovian quantum jumps method and a new 
type of unraveling which we call non-Markovian quantum diffusion. We combined 
the non-Markovian quantum jumps and non-Markovian quantum diffusion with kernel
smoothing techniques thus extending the applicability of these methods dramatically. 
Moreover, we also demonstrated the power of our approach with the {paradigmatic
amplitude damped driven two-level atom model}.
As an outlook, in addition of applying the methods for various
state-of-the art complex driven open quantum systems, it would be
interesting to investigate, e.g., what role the stochastic entropy
term in non-Markovian quantum diffusion plays in quantum stochastic
thermodynamics.
\bibliography{Diffusivelimit}
\end{document}